\documentclass[10pt,twocolumn]{article}
\usepackage{times}
\usepackage[utf8x]{inputenc}
\usepackage[final]{pdfpages}
\usepackage{t1enc}
\usepackage{graphicx}
\usepackage{textcomp}
\usepackage[T1]{fontenc}
\usepackage{mathabx}

\usepackage[square,comma,sort&compress,numbers]{natbib}
\usepackage[labelsep=period]{caption} 
\usepackage{mathabx}

\begin{document}

\twocolumn[
\centerline{\LARGE \bf Predicted dynamos for terrestrial extra-solar planets}
\centerline{\LARGE \bf  and their influence in habitability.} 

 \vskip10mm

 \begin{raggedright}
{\small {\bf Natalia G\'omez-P\'erez}, Mercedes L\'opez-Morales and Thomas Ruedas \\
Department of Terrestrial Magnetism, Carnegie Institution of Washington, 5241 Broad Branch Road, N.W., Washington DC 20015, USA. 
} 
\end{raggedright}

\vskip10mm %
]

\thispagestyle{empty}

%

\section{Introduction}

The question of habitability of terrestrial type extrasolar planets draws great interest.
The presence of a magnetospheric shielding to keep the planetary surface 
protected from the influence of high energy
particles 
appears to be key. 
Low mass stars have been proposed as good candidates to host habitable planets 
\cite{Tarter2007,Scalo2007}. However,  it has been stated  that their habitable zones
are at too short orbital distances to allow forplanetary magnetospheres strong enough to 
provide a shield from high energy particles \citep[e.g.,][]{Griessmeier2005}.

Around M-dwarfs (stars with masses less than 0.5 $M_{\Sun}$) the habitable zone
is restricted to orbital radii where a terrestrial planet in a circular orbit is expected to
be in synchronous rotation. 
If the orbital eccentricity is non-zero, 
the planet might end up in a stable non-synchronous rotational state
\cite{Correia2008}.
In either case, the rotation period of a close-in planet may be greatly diminished
by the tidal interaction with the star. 

Linear analysis of the magnetohydrodynamic equations have been 
expected to predict to first order the dipole magnetic moment of 
dynamos, although the fundamental magnetohydrodynamic
equations are highly non-linear \cite{Sano1993}.
Based on numerical models that take into account the whole set of
equations, it is also possible derive the magnetic dipole moment of 
dynamo in rotating spheres and predict what are the important parameters
that determine the magnitude and geometry of the planetary dynamo field 
\cite{Olson2006}.

In this paper, we find the internal stratification for terrestrial planets with  given mass-radius pairs, 
and use the core size and density to estimate their maximum dipolar magnetic moment. 
We also comment on the temporal evolution, 
although more information (e.g.,  core composition,
mantle rheology and history)
is crucial in determining the state of the dynamo with planetary age.

\section{Methodology}

We choose to model a planet with two layers, i.e., a rocky mantle and an iron core.
Given the radius ($R$) and mass ($M$) of the planet, 
we integrate density and pressure as a function of radius, 
assuming adiabatic compression. 
The integration is performed from the surface to the center of the planet using:
\begin{center}
\begin{tabular}{cc}

\begin{minipage}{\columnwidth*5/11}

\begin{equation}
\frac{d\rho}{dr}=\frac{\rho\,g}{\phi},
\end{equation}
\end{minipage}

&

\begin{minipage}{\columnwidth*5/11}

\begin{equation}
\frac{dP}{dr}=-\rho\,g,
\end{equation}
\end{minipage}
%
%

\end{tabular}
\begin{minipage}{\columnwidth*5/11}
\begin{equation}
\frac{dm}{dr}=4\,\pi\,\rho\,r^2,
\end{equation}
\end{minipage}
\end{center}
where $\rho$ is the radially dependent density, 
$r$ is the planetary radius,
$g$ is the gravitational acceleration,
$\phi$ is the seismic parameter,
$P$ is the pressure, and $m(r)$ is the mass in a sphere of radius $r$.

The seismic parameter, $\phi$, is calculated using the third order Birch-Muranham finite strain equation \citep[e.g.,][]{Jackson1998},
\begin{equation}
\phi(r)=\frac{1-2\varepsilon}{3\rho_0}  \left(C_1\, (1- 7 \varepsilon) + C_2 (\varepsilon - 9/2\, \varepsilon ^2)\right),
\end{equation}
where 
$\varepsilon = 1/2 \,\left( 1-  (\rho(r)/\rho_0)^{2/3} \right)$, $C_1=3\,K_{S0}$,  and $C_2=9\,K_{S0}\,( 4-K'_{S0} )$.
The subscript zero indicates surface pressure and temperature (i.e., $P=0$ $T=$300K). 
$K_{S}$ and $K'_{S}$ are the adiabatic bulk modulus and its first order derivative with respect to pressure, respectively.
Using a fifth-order Runge-Kutta integration, 
we solve for the total mass to be within 1\% of the given planetary mass, $M$. 

We integrate the equation for one single type planet with a mantle composition comparable
to that of the Earth's lower mantle, and a pure iron core (i.e., $K_{S0} = 130$GPa,  $K'_{S0} = 4.27$, and $\rho_0 = 3350\, \mathrm{kg \,m}^{-3}$ for the mantle and
$K_{S0} =160.2$GPa,  
$K'_{S0} = 5.82$,  and $\rho_0 = 8300\, \mathrm{kg \,m}^{-3}$ for the core \cite{Valencia2007}).

\section{Results}

\begin{figure}[h]
\centerline{\includegraphics[width=\columnwidth]{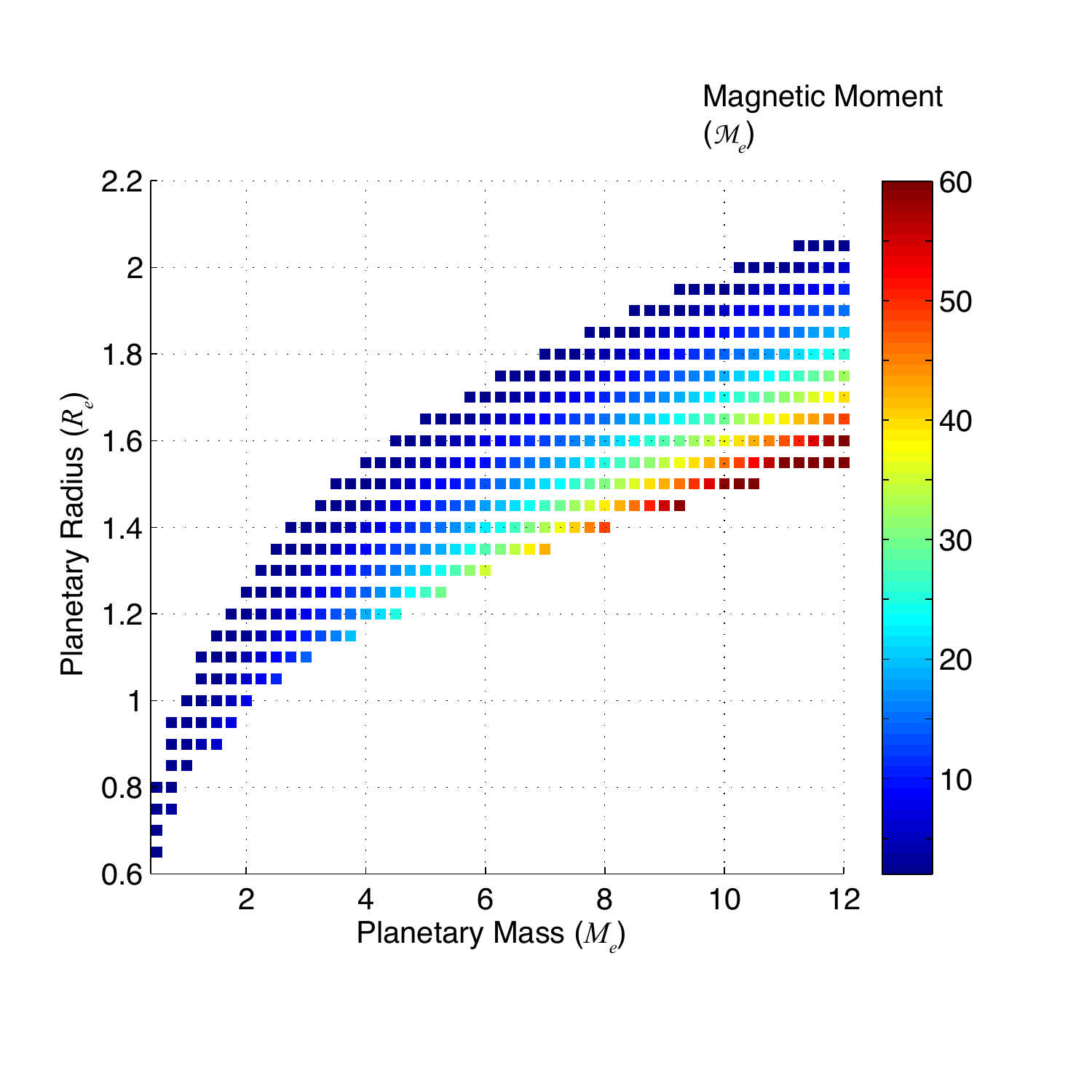}}
\caption{Dipole magnetic moment in units of $\mathcal{M}_e$ for a given planetary mass, 
shown units of Earth $R_e$, and radius, in units of Earth masses $M_e$.}
\label{fig:MagMom}
\end{figure}

We calculate the magnetic dipole moment as function of planetary mass and radius
for our simple model, by using the dynamo field scaling law \cite{Olson2006}
\begin{equation}
\mathcal{M}\simeq4\pi\, r_o^3\, \gamma \,(\bar{\rho_o}\mu_0)^{1/2}(FD)^{1/3}, \label{eq:MagMom}
\end{equation}
where $\mathcal{M}$ is the magnetic dipole moment of the dipole-dominated field,
$r_o$ is the radius of the iron core,
$\gamma$ is a fitting scaling parameter with a value 0.2, 
$\bar{\rho_o}$ is the iron core bulk density,
$\mu_o$ is the magnetic permeability of vacuum,
$F$ is the convective heat flux, assumed to scale as $r_o^2$,
and $D$ is the thickness of the shell encasing the liquid core.
Here we use $D=0.65r_o$ for which the strongest dipolar magnetic field is expected \cite{Heimpel2005a}.
We show our estimate of $\mathcal{M}$ in figure~\ref{fig:MagMom}.

$(R,M)$ pairs that fall above the top envelope do likely have substantial liquid oceans.
Pairs falling below the lower envelope correspond to compositions 
of a material denser than pure iron, so that region is forbidden.

\section{Conclusions}

As indicated in equation~\ref{eq:MagMom} the predicted dipole moment 
depends on the bulk density of the fluid core. For pure iron cores,
the bulk density depends only on the pressure at the core-mantle boundary.
The iron self compression yields a density variation from  0.98-1.91 that of Earth's fluid core
giving a variation of a factor of $\sqrt{2}$ in $\mathcal{M}$. 
A more significant effect is expected from the size of the core, $r_o$, which is found to
vary between 0.2-2.8 that of Earth, which can increase $\mathcal{M}$ by a factor of up to 60.
Light elements are likely present in planetary cores. 
If this is the case,  the size of the core would take a larger fraction of the 
total radius and its density
would be  lower. Upper and lower envelopes
move further up in radius but  the range of $\mathcal{M}$ 
for the same calculation with a metallic core composition enriched with light element, 
i.e.,  $\mathrm{Fe}_{0.8}\mathrm{(FeS)}_{0.2}$,
is comparable to that of pure iron.

%

%
We predict that planets close-in to the star may have significant magnetic fields that allow 
a magnetosphere to be present. For planets developing a magnetic field under the influence of a strong, 
time-invariant external magnetic field, modifications from the usual dynamo scaling is likely \cite{Gomez2010b},
but the presence of a substantial magnetic field  (i.e., able to form a magnetosphere)
cannot be ruled out.

\bibliography{MMExoterrestrial}
\bibliographystyle{LPSCshort_url}

\end{document}